\documentclass[doublespacing]{elsart}
\usepackage{natbib}
\usepackage{epsfig}
\usepackage{graphicx}
\usepackage{dcolumn}
\usepackage{bm}

\begin{document}

\begin{frontmatter}
\title{Generalized Einstein Relation in an aging colloidal glass}
\author{B\'ereng\`ere Abou, Fran\c cois Gallet, Pascal Monceau \and
  No\"elle Pottier}
\address{Laboratoire Mati\`ere et Syst\`emes Complexes (MSC), UMR CNRS
  7057\\ Universit\'e Paris 7-Denis Diderot\\ 10, rue Alice Domon et
  L\'eonie Duquet\\ 75205
  Paris Cedex 13, France}
\date{Received : date / Revised version : date}
\begin{abstract}

We present an experimental and theoretical investigation of
the Generalized Einstein Relation (GER), a particular form of a
fluctuation-dissipation relation, in an out-of-equilibrium visco-elastic
fluid. Micrometer beads, used as thermometers, are immersed in an
aging colloidal glass to provide both fluctuation and dissipation
measurements. The deviations from the Generalized Einstein Relation are derived
as a function of frequency and aging time. The observed
deviations from GER are interpreted as directly related to the change in the glass relaxation
times with aging time. In our scenario, deviations are observed in the regime where the
observation time scale is of the order of a characteristic relaxation
time of the glass.

\end{abstract}

\begin{keyword}

\PACS 64.70.Pf \sep 05.40.-a \sep 05.70.-a
\end{keyword}
\end{frontmatter}
%
\section{Motivation}
During the past years, slow relaxation systems have received
considerable attention in the sense that they constitute a challenge
for new developments in nonequilibrium statistical mechanics. Such systems, which are rather common in nature, include
structural glasses, polymers, colloids and granular matter. They share
some characteristics called {\em glassy behavior}, in particular a
drastic slowing down of relaxation processes when some control
parameters are varied. As the characteristic relaxation times becomes
longer than, or comparable to, the observation timescale, the system
is said to age. The physical properties of the material depend on the
waiting time, which is the time elapsed since preparation in the
nonequilibrium state.

Recently, there have been serious attempts to approach and understand
aging phenomena on a theoretical level. Many efforts have been devoted
to apply statistical physics description -- in particular,
fluctuation-dissipation relations (FDR) -- to out-of-equilibrium
systems. At thermodynamic equilibrium, fluctuation-dissipation
relations relate the response functions of the system to its
autocorrelation functions, involving a single thermodynamic parameter,
the equilibrium temperature. However, FDR are not expected to hold in
out-of-equilibrium systems. The idea that a timescale dependent
nonequilibrium or {\em effective} temperature could describe the slow
relaxation modes has received a lot of attention. This effective
temperature, defined from an extension of FDR, is different from the
bath temperature and has been shown to display many of the properties
of a thermodynamic temperature \cite{cug93,cug97}. Deviations from FDR
have been observed in many numerical simulations
\cite{parisi,barrat,sellitto,marinari,barrat2,barrat3,makse,fielding}. To
date, there are still few experiments where FDR are studied in aging
materials, such as structural glasses \cite{grigera}, colloidal
glasses \cite{bellon,bellon2,prl93,strachan,jabbari,greinert}, spin glasses
\cite{herisson}, polymer glasses \cite{buisson} and granular media
\cite{danna,oja}. In most cases, they show deviations from FDR as a function of
waiting time and frequency. Unfortunately, no global understanding can
yet be extracted from these experimental findings.

Here we present an experimental and theoretical
investigation of the Generalized Einstein Relation (GER), a particular
form of a fluctuation-dissipation relation, in an out-of-equilibrium
visco-elastic fluid. The deviations from the GER are measured as a
function of waiting time $t_w$ and frequency $\omega$, leading to an
effective temperature $T_{\mbox{\scriptsize eff}}(\omega)$,
parametrized by $t_w$. This is achieved by simultaneously
measuring the response function to an external force -- using an
optical tweezer -- and the position fluctuations of micrometric beads
embedded in the glass.

The
paper is organized as follows. 
Sec. \ref{theory} is devoted to the theoretical investigation of the Generalized
Einstein Relation in an out-of-equilibrium system, leading to the
definition of an effective temperature. Sec. \ref{experiment} presents
the experimental procedures used to test such a GER in an aging
colloidal glass: Brownian motion measurements and response to an oscillatory force -- using optical tweezer. In
Sec. \ref{results}, the effective temperature is derived as a function of aging
time and frequency. In Sec. \ref{discussion}, we suggest a microscopic interpretation of
our findings. We conclude the paper
with comparisons with other tests of FDR in the same glass and with open
questions. 

\section{Fluctuation-dissipation relation in an out-of-equilibrium environment}
\label{theory}

\subsection{Equilibrium medium case}

Let us first consider a diffusing particle of mass $m$ evolving in a stationary medium. Its motion can be described by a generalized Langevin equation, in which $v(t) = dx/dt$ is the particle velocity, $F(t)$ the random force acting on the particle, and $\gamma (t)$ a delayed friction kernel that takes into account the visco-elastic properties of the medium :
\begin{equation}
m\,{dv\over dt}+m\int_{-\infty}^\infty\gamma(t-t')v(t')\,dt'= F(t)
\label{GLE}
\end{equation}
Note for further purpose that, in the usual experimental frequency range, the inertial term in Eq.(\ref{GLE}) is quite negligible. The frequency-dependent particle mobility thus reduces to $\tilde\mu(\omega)=[m\tilde\gamma(\omega)]^{-1}$, where $\tilde\gamma(\omega)=\int_{-\infty}^\infty\gamma(t)e^{i\omega t}\,dt$ is the Fourier transform of the delayed kernel $\gamma(t)$.

\subsubsection{The Generalized Einstein Relation}

If the surrounding stationary medium is in thermal equilibrium at
temperature $T$, one can write a Kubo formula expressing the
frequency-dependent particle mobility $\tilde\mu(\omega)$ in terms of
the velocity correlation function \cite{Kubo} :
\begin{equation}
\tilde\mu(\omega)={1\over kT}\int_0^\infty\langle v(t)v(t=0) \rangle e^{i\omega t}dt
\label{Kubo}
\end{equation}
(the symbol $\langle\ldots\rangle$ denotes the average over an ensemble of realizations). Formula (\ref{Kubo}) defines $\tilde\mu(\omega)$ as an analytic function of $\omega$ in the upper complex half plane. 

From Eq. (\ref{Kubo}), one can derive a formula linking the drift and
diffusion properties of the particle. The mean-squared displacement of
the diffusing particle, as defined by :
\begin{equation}
\langle{\Delta x^2}(t)\rangle=\langle{[x(t)-x(t=0)]}^2\rangle,\qquad t>0
\label{displacement}
\end{equation}
can be deduced from the velocity correlation function {\it via\/} a double integration over time:
\begin{equation}
\langle{\Delta x^2}(t)\rangle=2\int_0^tdt_1\int_0^{t_1}dt_2\langle v(t_1)v(t_2)\rangle
\end{equation}
Introducing the Laplace transformed quantities $\hat v(s)=\int_0^\infty v(t)e^{-st}dt$ and $\langle{\Delta x^2}(s)\rangle=\int_0^\infty\langle\widehat{\Delta x^2}(t)\rangle e^{-st}dt$, one gets, by Laplace transforming 
Eq. (\ref{displacement}):
\begin{equation}
s^2\langle\widehat{\Delta x^2}(s)\rangle=2\,\langle\hat v(s)v(t=0)\rangle
\label{Laplace deplacement}
\end{equation}
The quantity $\langle\hat v(s)v(t=0)\rangle$ can be obtained by setting $\omega=is$ in the Kubo formula 
({\ref{Kubo}) (which is allowed for positive $s$, since the function
  $\tilde\mu(\omega)$ is analytic in the upper complex half
  plane). One gets :
\begin{equation}
\langle\hat v(s)v(t=0)\rangle=kT\,\hat\mu(s)
\label{vv}
\end{equation}
with $\hat\mu(s)=\tilde\mu(\omega=is)$. Eq. (\ref{Laplace
  deplacement}) then takes the form of a relation linking the Laplace
transforms of the particle mean-squared displacement and mobility : 
\begin{equation}
s^2\langle\widehat{\Delta x^2}(s)\rangle = 2kT\,\hat{\mu}(s) 
\label{GER}
\end{equation}
Eq. (\ref{GER}) is known as the Generalized Einstein Relation (GER).

\subsubsection{The fluctuation-dissipation relation}

It follows from Eq. (\ref{Kubo}) that the velocity spectral density is related to the dissipative part of the mobility by a fluctuation-dissipation relation (namely, the celebrated Einstein relation):
\begin{equation}
C_{vv}(\omega)=\int_{-\infty}^\infty\langle v(t)v(t=0)\rangle e^{i\omega
  t}\,dt=2kT\Re e\tilde\mu(\omega)
\label{Einstein}
\end{equation}
The validity of Eq. (\ref{Einstein}) is restricted to $\omega$ real. 

\subsubsection{Discussion}

The Einstein relation (\ref{Einstein}) contains the same information
as the GER (\ref{GER}), since $\tilde\mu(\omega)$ can be deduced from
$\Re e \tilde\mu(\omega)$ with the help of the usual Kramers-Kronig relations valid for real $\omega$. Thus, Eqs. (\ref{GER}) and (\ref{Einstein}) constitute fully equivalent formulas, which both involve the thermodynamic bath temperature $T$. 

\subsection{Out-of-equilibrium environment}

The general situation of a particle diffusing in an out-of-equilibrium environment is much more difficult to describe. As well-known, in an aging medium, no well defined thermodynamical temperature does exist, so  that Eqs. (\ref{GER}) 
and (\ref{Einstein}) are no longer expected to be valid. 

We address the question whether the study of the diffusion and drift of the probe particle is likely to provide information about the out-of-equilibrium properties of its surrounding medium.

\subsubsection{Modified Generalized Einstein Relation}

Out of equilibrium, one can write a modified Kubo formula as :
\begin{equation}
\tilde\mu(\omega)={1\over k\tilde\Theta(\omega)}\int_0^\infty\langle v(t)v(t=0)\rangle e^{i\omega t}\,dt
\label{modKubo}
\end{equation}
Formula (\ref{modKubo}) defines the product $\tilde\mu(\omega)\tilde\Theta(\omega)$ as an analytic function of $\omega$ in the upper complex half plane. Since this analyticity property holds for $\tilde\mu(\omega)$, it also holds for $\tilde\Theta(\omega)$.
 
Accordingly,  Eq. (\ref{vv}) has to be replaced by :
\begin{equation}
\langle\hat v(s)v(t=0)\rangle=k\hat\Theta(s)\hat\mu(s)
\end{equation}
Thus, the relation linking the Laplace transforms of the particle
mean-squared displacement and mobility writes :
\begin{equation}
s^2\langle\widehat{\Delta x^2}(s)\rangle = 2k\hat\Theta(s)\hat{\mu}(s) 
\label{modGER}
\end{equation}
Eq. (\ref{modGER}) states an out-of-equilibrium GER. It involves a $s$-dependent GER ratio $\hat\Theta(s)$, parametrized by the age of the system $t_w$.
 
Experimentally, it is always possible to measure {\em independently}
both quantities -- $\langle \widehat{\Delta x^2} (s) \rangle$ and
$\hat{\mu}(s)$ -- during aging. For a particle evolving in a bath at
equilibrium, $\hat{\Theta}(s)= T$ would be independent on $s$, as
shown by Eq. (\ref{GER}). In an out-of-equilibrium environment where
$\hat\Theta(s)$ is {\em a priori} expected to be different from $T$, the mobility of the probe particle cannot be
deduced from the mean-squared displacement measurement, and {\it
vice-versa}. As a consequence, the visco-elastic properties of the
out-of-equilibrium medium cannot be deduced from the single
measurement of the probe thermal fluctuations and passive
microrheology cannot be achieved [21].

\subsubsection{Modified fluctuation-dissipation relation}

Since, in an out-of-equilibrium environment, even stationary, the
Einstein relation (\ref{Einstein}) is not {\em a priori} satisfied, it has been
proposed \cite{cug93,cug97}  to rewrite it in a modified way with the help of a frequency-dependent effective temperature. Such a
quantity, denoted as $ T_{\rm eff} (\omega)$ and parametrized by the
age of the system $t_w$, has been defined, for real $\omega$, {\em
  via\/} an extension of the Einstein relation (\ref{Einstein}). One
writes a modified Einstein relation as \cite{pottier2003,pottier2005} :
\begin{equation}
C_{vv}(\omega)= 2kT_{\rm eff}(\omega)\Re e\tilde\mu(\omega)
\label{modEinstein1}
\end{equation}
It has been argued in \cite{cug93,cug97} that the effective temperature $T_{\rm eff}(\omega)$ defined in such a way would have a possible thermodynamic meaning, in the sense that it plays the same role as the thermodynamic temperature in systems at equilibrium (namely, it controls the direction of heat flow and acts as a criterion for
thermalisation).

\subsubsection{Discussion}

In the case of an out-of-equilibrium medium, one has at hand the
modified GER (\ref{modGER}) and the modified FDR
({\ref{modEinstein1}). The question thus arises of the link between
these two descriptions, namely of the relation between the GER ratio
$\hat\Theta(s)$ and the effective temperature $T_{\rm eff}(\omega)$.

The modified Kubo formula (\ref{modKubo}) defines a complex
fre\-quency-dependent function $\tilde\Theta(\omega)$, in terms of which
the velocity spectral density writes:
\begin{equation}
C_{vv}(\omega)=2k\Re e \bigl[\tilde\Theta(\omega)\tilde\mu(\omega)\bigr]
\label{modEinstein2}
\end{equation}
Comparing Eqs. (\ref{modEinstein1}) and (\ref{modEinstein2}), one gets
the following relationship between $\tilde\Theta(\omega)$ and $ T_{\rm
eff}(\omega)$ :
\begin{equation}
T_{\rm eff}(\omega)\,\Re e \tilde\mu(\omega)=\Re
e\bigl[\tilde\mu(\omega)\tilde\Theta(\omega)\bigr]
\label{Theta and Teff}
\end{equation}
Eq. (\ref{Theta and Teff}) displays the fact that the effective temperature $T_{\rm eff}(\omega)$ involved in the modified Einstein relation (\ref{modEinstein1}) can be deduced from the GER ratio introduced in the modified GER (\ref{modGER}) (and {\em vice-versa\/}). This point has been developed in details in \cite{pottier2005}. 

Thus, as it can be seen from Eq. (\ref{modGER}), independent
measurements of the particle mean-squared displacement and mobility in
an aging medium give access, once $\langle\widehat\Delta
x^2(s)\rangle$ and $\hat\mu(s)=\tilde\mu(\omega=is)$ are determined, to the
GER ratio $\hat\Theta(s)$. Eq. (\ref{Theta and Teff}) then allows to
derive an effective temperature $T_{\rm eff}(\omega)$ from the
experimentally measurable GER ratio
$\tilde\Theta(\omega)=\hat\Theta(s=-i\omega)$.

In the out-of-equilibrium case, the modified GER (\ref{modGER}) and the modified FDR (\ref{modEinstein1}) constitute fully equivalent formulas. However, Eq. (\ref{modGER}) involves a GER ratio $\hat\Theta(s)$, naturally expressed in terms of the Laplace variable $s$, while Eq. (\ref{modEinstein1}) involves an effective temperature $T_{\rm eff}(\omega)$, naturally expressed in terms of the real frequency $\omega$.
  
\section{Experimental procedures}
\label{experiment}

\subsection{Samples preparation}
The experiments were performed on Laponite RD, a synthetic clay
manufactured by Laporte Industry. The particles of Laponite are
colloidal disks of 25 nm diameter and 1 nm thickness, with a negative
surface charge on both faces \cite{Kroon}. The clay powder was mixed
in ultra-pure water, and the pH value of the suspensions fixed to pH =
10 by addition of NaOH, providing chemically stable particles
\cite{Thompson and Butterworth}. The suspension was stirred vigorously
during 15 minutes and then filtered through a Millipore Millex - AA
0.8 $\mu$m filter unit. This procedure allows us to prepare 
a reproducible initial liquid state. The aging time $t_w = 0$ of the
suspension is defined as the moment it passes through the filter.

These aqueous suspensions form glasses for low volume fraction in
particles \cite{wigner}. Starting from a viscous ``liquid'' state right after
preparation, the suspension becomes more and more visco-elastic with
time. Since the physical properties of the suspension depend on the
time $t_w$ elapsed since preparation, the sample is said to age. Aging
can be seen through the change in both the visco-elastic properties
and of the colloidal disks diffusion
\cite{aboupre64,jorlapo}. Laponite suspensions age on timescales that
depend on the particles' concentration. We are thus able to control the
aging timescales of the glass by adjusting this concentration. With a
volume fraction of $2.3 \% $ wt, the glass evolves over several hours, slowly enough to
allow us quasi-simultaneous measurements of the fluctuation and dissipation properties, within a few minutes, without significative aging of the sample. These two successive measurements are
thus considered to be performed at the same waiting time $t_w$.

The experiments were carried out in a square chamber -- $20 \times 20$
mm$^2 $ -- made of a microscope plate and a coverslip separated by a thin
spacer ($0.1$ mm thickness). The beads are suspended in the glass right after its preparation. The chamber is then filled with the
suspension, sealed with vacuum grease and mounted on a piezoelectric
stage on the optical microscope plate. The probes are latex and
silica beads, in very low concentration (respectively $10^{-4} \%$ and
$4.10^{-4}  \%$ in volume). Latex beads ($1.0 \pm 0.1 \mu$m in diameter,
Polysciences, Inc.), were preferentially used for fluctuation
measurements : since they do not sediment during the experiment, their
random motion is not perturbed by the chamber walls. Silica beads
($2.1 \pm 0.1 \mu$m in diameter, Bangs Lab Inc.) were used for dissipation
measurements, because they are more efficiently trapped by the optical
tweezers. By assuming that the particle mobility scales
as the inverse particle size according to $\tilde \mu(\omega) = 1/ 6 \pi
R \tilde \eta (\omega)$ (Stokes assumption in a visco-elastic
fluid), the comparison between the results of the fluctuation and
dissipation measurements, once rescaled to the same diameter, is
meaningful. This relation is exact in a purely viscous fluid at low
Reynolds number; here we
assume its validity in the visco-elastic fluid.

In these experiments, the nature of the bead material may play a role through
the particle-sample interactions, dominated by
electrostatics. Actually, the suspensions of Laponite are prepared at
pH = 10 by addition of NaOH. At this ionic concentration, the
Debye-H\"uckel screening length is of the order of $30$ nm, which is
comparable to the Laponite particle diameter, and smaller than the
interparticle distance. In such conditions, the electrostatic
interactions between the micro beads (actually both silica and latex
beads are charged) and the Laponite particles are screened on a
distance much smaller than the bead diameter. As a consequence, the
range of interactions between beads and particles does not exceed $30$
nm and the system is not perturbed on large scales by the introduction
of the beads.

\subsection{Thermal fluctuations of the beads}
At a given aging time $t_w$, we record the fluctuating motion of 1
$\mu$m latex beads during 8 s, with a fast CCD camera sampling at 250
Hz (Fastcam-PHOTRON LTD). A digital image analysis allows us to track the
bead positions $x(t)$ and $y(t)$ close to the focus plane of the
microscope objective. For each bead, we calculate the time-averaged
mean-squared displacement $ \langle \Delta r^2 (t) \rangle _{t'} =
\langle[ x(t'+ t) - x(t') ]^2 + [y(t' + t) - y(t') ]^2 \rangle_{t'} =
2 \langle \Delta x^2 (t) \rangle_{t'}$. To preserve a good statistics,
we keep the data of $\langle\Delta r^2 (t)\rangle_{t'}$ in the range
$0.004 < t < 1$ s. The resolution on the bead
position, determined by sub-pixel accuracy in the image analysis
detection, is $10$
nm. The glass remains in a quasi-stationary state
during the recording, which takes a short time compared to the aging
timescale. The quantity $\langle\Delta r^2 (t)\rangle_{t'}$, averaged
over several beads and realizations, can thus be identified to the
ensemble-averaged mean-squared displacement.

\subsection{Dissipative response to an oscillatory force}
We describe now the measurement of the mobility $\tilde \mu(\omega)$,
at a given frequency $\omega $, for various aging times $t_w$. This
measurement is performed immediately after the Brownian motion
recording, at the same aging time $t_w$. Since the aging Laponite
suspension is a visco-elastic fluid, the bead mobility $\tilde
\mu(\omega) = |\tilde \mu(\omega)| e^{i \varphi(\omega)} $ is a
complex number. We thus need to measure the phase and modulus of the
tracer mobility. We use an optical tweezer to trap a 2.1 $\mu$m silica
bead immersed in the glass. Trapping is achieved by focusing a
powerful infrared laser beam (Nd YAG, Spectra-Physics, $P_{max} = 600$
mW) through a microscope objective of large numerical aperture
\cite{henon}. The trapping force $F$ on a small dielectric object like
a silica bead is proportional to the intensity gradient in the
focusing region. It depends on the distance $x$ of the center of the trapped object
to the center of the trap, according to $F(x)=-kx(1+\epsilon|x|)$. The
corrective factor $\epsilon$ is introduced to take into account the
non-harmonicity of the trapping potential. The trap stiffness $k$ and the factor $\epsilon$
are determined from an independent calibration. The calibration
procedure was described in details in previous publications using
optical tweezers set-up \cite{balland}. Once the bead is
trapped, we make the experimental chamber oscillate by monitoring the
displacement $x_p \exp(-i \omega t)$ of a piezoelectric
stage. Neglecting in a first step the visco-elastic fluid inertia, the
fluid displacement $x_f$ in the bead vicinity is taken equal to the
piezolectric stage displacement $x_p$. The validity of this assumption
is discussed in the Appendix. Due to the relative
bead/fluid motion, the visco-elastic fluid exerts a sinusoidal force
$F' \exp(- i \omega t)$ on the bead. We record with the fast camera the
bead movement, and measure by conventional image analysis its
displacement $x \exp{(-i \omega t)}$ from the trap center. Notice that
$x$ is a complex number which includes a phase shift due to the fluid
visco-elasticity. At a given frequency $\omega$, the force amplitude
$F'(\omega)$ is given by $\tilde F'(\omega) = \tilde v(\omega)/
\tilde\mu (\omega)$, where $\tilde v(\omega) = i \omega (x_f - x)= i
\omega (x_p - x)$ is the relative chamber wall / bead velocity, and $\tilde
\mu(\omega)$ the Fourier transform of the bead mobility. In our range
of experimental frequencies ($0.5 \leq f \leq 10$ Hz), the bead
inertia is negligible, so that we can simply use the relation $F + F'
= 0$ to calculate $|\tilde \mu(\omega)|$ and $\varphi(\omega)$. Notice
that the motion of the piezoelectric stage is numerically controlled
by a sequence of successive sinusoidal signals at five different
frequencies $\{0.5, 1, 2, 5, 10 \}$ Hz. The same program synchronously
generates a sequence of pulses to trigger the image acquisition, so
that the phase shift between the force and the bead movement can be
accurately measured.
For both the passive and the active
microrheology experiments, we used a conventional image processing
software (IMAQ vision builder from National Instruments) to determine
the bead position from the numerical recording of the bead images. The
accuracy on the bead position was about $10$ nm, for typical motions
comparable to the bead radius, in the micrometer range.

\section{Results}
\label{results}

\subsection{Rescaling of the aging dynamics}
\label{scaling}

Figure \ref{scale}(a) shows the increase in the viscosity modulus with waiting
time for a set of six different realizations. The complex mobility of a Laponite
suspension at $2.3 \%$ wt was measured by applying an oscillatory
force on a silica bead in the linear regime. The complex viscosity
modulus $|\tilde{\eta} (\omega)|$ was estimated from the complex
mobility modulus $|\tilde{\mu} (\omega)| $ assuming that the Stokes
relation $\tilde{\mu} (\omega)= \frac{1}{6\pi R
\tilde{\eta} (\omega)}$ remains valid in the visco-elastic fluid. Starting from a
value close to the water viscosity $10^{-3}$ Pa.s, the glass viscosity modulus is shown to increase by $3$ orders of magnitude over about $200-400$
minutes. These variations in aging timescales are not so surprising
because in such suspensions, the aging dynamics is known to
drastically depend on the particle concentration. In a suspension at $2.5 \%$
wt, a three orders of magnitude increase of the complex viscosity
modulus -- measured with a conventional rheometer -- is reached after $100$ minutes, while in a $3.5 \%$ wt suspension,
the same increase is reached after $10$ minutes
\cite{jorlapo}. Besides, slight ionic concentration variations in the suspension induce important changes in
particle interactions. These variations may arise either from pH
differences or ions salting-out from the chamber walls. 

Here we suggest a time rescaling in order to compare the different
realizations. We assume that the differences
in the aging dynamic shown in Fig. 1(a) can be compensated by a linear
stretching in aging time. This is supported by a recent paper by Joshi
where the glass transition of a Laponite suspension is investigated
\cite{joshi}. The characteristic dimensionless relaxation time $\tau /
\tau_0 $ of the suspension
before the system enters the full aging regime ($t_w < {t_w}^{*}$) is shown to only depend on the
dimensionless age $ t_w / {t_w}^{*}$. The transition time ${t_w}^{*}$ depends on
the volume fraction $\Phi$, according to $t_w^{*}\sim
\Phi^{-2/3}$. Given a massic
fraction of 2.3 $\%$, and the aging timescales involved, our
experiments were performed before the full aging regime takes place \cite{aboupre64}. By assuming that i) the suspension viscosity only
depends on the characteristic relaxation time $\tau/ \tau_0$ and thus
on $t_w / {t_w}^{*}$ \cite{prl89}, and ii) that the differences between realizations
are essentially due to slight volume fraction variations, it seems
realistic to balance the differences by a linear stretching in aging
time. In Fig. \ref{scale}(b), the aging time was linearly stretched to
make the viscosities roughly coincide for all experiments, choosing a
particular realization as a reference. Nevertheless, it must be clear
that the same conclusions on $T_{\rm eff}$ are obtained on each
separate sample.  With this rescaling, we are now able to average over
different realizations.  From now on, the aging times of the different
realizations are rescaled by using this rescaling.
 \begin{figure}
\includegraphics[width=7cm,clip]{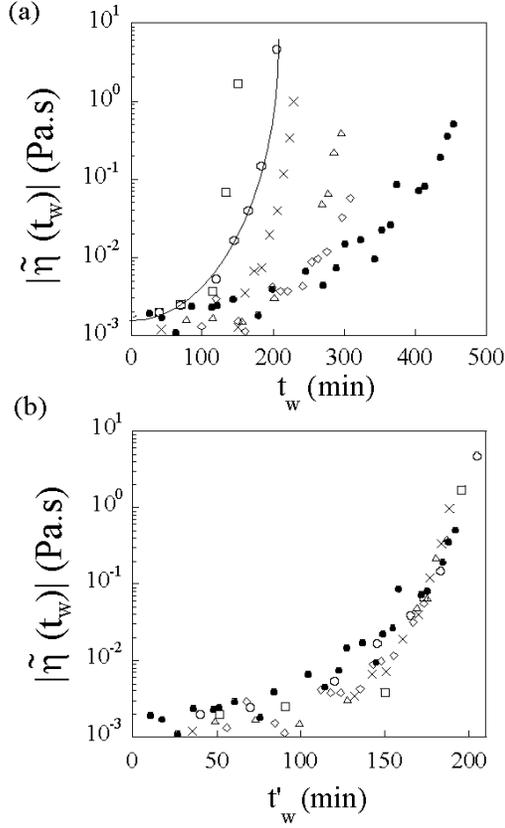}
\caption{(a): Viscosity modulus of the glass as a function of aging
  time, for six different realizations. The viscosity modulus was
  measured by applying an oscillatory force at a frequency $f = 1 $ Hz
  on a silica bead in the linear regime; (b): Rescaling of the complex
  viscosity modulus. The aging time is linearly stretched $t_w \to
  t'_w=\alpha t_w$ to make the
  viscosities roughly coincide for all experiments, choosing a
  particular realization as a reference.}
\label{scale}
\end{figure}

\subsection{Brownian motion of beads in the glass}
Figure \ref{brown} shows the mean-squared displacement of latex
beads immersed in the colloidal glass, as a function of time $t$, for
various aging times $t_w$. At $t_w=0$, we observe a nearly diffusive
behavior of the tracer beads, characterized by a linear dependency of
the mean-squared displacement with time. Upon increasing on $t_w$, the
tracer motion becomes sub-diffusive.
\begin{figure}[h]
\includegraphics[width=8cm,clip]{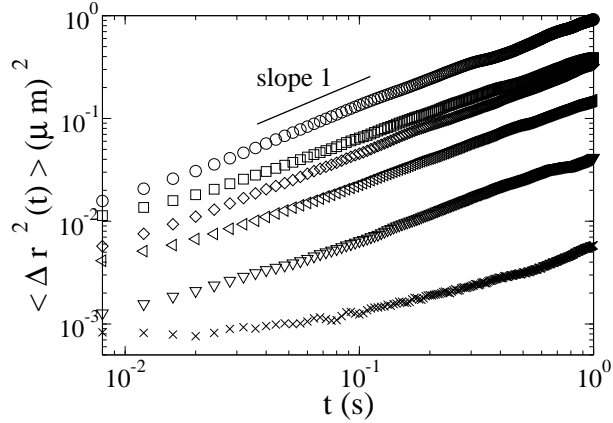}
\caption{Mean-squared displacement of 1 $\mu$m
latex beads immersed in the glass, as a function of
time. The curves correspond to different aging times $t_w =
15, 64, 100, 148, 169$ and $186$ minutes from top to bottom, measured
for one realization. The
fluctuating motion is purely diffusive at short $t_w$, and
becomes sub-diffusive as the glass ages.}
\label{brown}
\end{figure}

\subsection{Mobility from oscillatory force}

Figure \ref{mu} shows the frequency and aging time dependence of the
complex mobility modulus $|\tilde \mu(\omega)|$ and phase
$\varphi(\omega)$ measured when applying an oscillatory external force
at successive frequencies $\{0.5, 1, 2, 5, 10\}$ Hz. The mobility
modulus $ |\tilde\mu(\omega)|$ of the bead was found to decrease with
aging time $t_w$ as shown in Fig. \ref{mu}(a). This corresponds to the
increase in the glass visco-elastic modulus. Fig. \ref{mu}(b)
represents the phase shift $\varphi(\omega)$ as a
function of $t_w$. Starting from zero, the phase $\varphi$ decreases
with $t_w$. This corresponds to the shift between a purely viscous
behaviour ($\varphi=0$) to a visco-elastic one ($-\pi/2 < \varphi < 0$). In a
first-order approximation, the phase remains independent of the frequency.
Figure \ref{mu}(c) shows the same $ |\tilde\mu(\omega)|$ data, plotted
as a function of the frequency $\omega$, for different aging times
$t_w$. The mobility modulus $|\tilde \mu(\omega)|$ is well fitted by a
power law $|\tilde \mu(\omega)| = \mu_0 (\omega/\omega_0)^{\beta}$ in
the experimental frequency range corresponding to one and half
decade. Here, $\omega_0$ is an arbitrary reference frequency set to
$\omega_0 =1$ rad.s$^{-1}$ for convenience. Starting from about zero at
low aging times, the exponent $\beta$ increases with $t_w$. Since $
|\tilde\mu(\omega)|$ exhibits a power-law behavior, the dependence of
$\varphi(t_w)$ must be related to $\beta$ by $\varphi (t_w)= -
\beta(t_w) \pi/2$. This ensures that $\hat \mu (s)$ is a real number. In
Fig. \ref{mu}(b), we plot the quantity $- \beta(t_w) \pi/2$ (full
circles), which is in fair agreement with the experimental data
$\varphi (t_w)$.
\begin{figure}[h]
\includegraphics[width=7.5cm,clip]{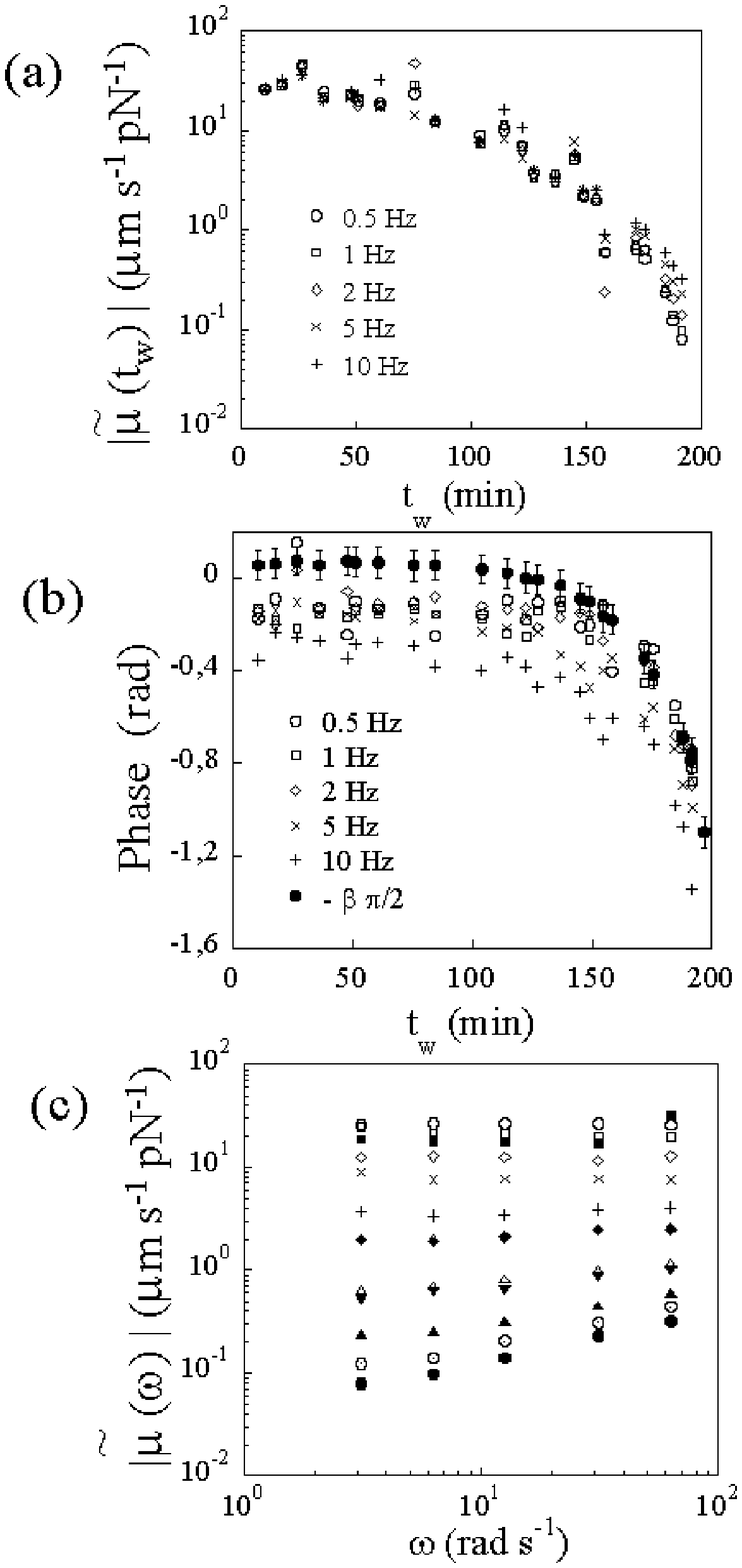}
\caption{Modulus (a) and phase (b) of the complex mobility
  $\tilde\mu(\omega)= |\tilde\mu(\omega)| e^ {i\varphi(\omega)}$ of
  the bead as a function of $t_w$, for various frequencies of the
  applied force; (c) mobility modulus versus $\omega$ for various
  $t_w$; from top to bottom, $t_w$ varies from $8$ to $191$ min, for
  this realization. At low $t_w$, $|\tilde\mu(\omega)|$ is nearly independent of
  $\omega$. Upon increasing $t_w$, the modulus is well fitted by a
  power-law $|\tilde \mu(\omega)|=\mu_0 (\omega/\omega_0)^{\beta}$ with
  $\omega_0=1$ rad.s$^{-1}$ and $\beta$ only
  depending on $t_w$. The full circles in (b) correspond to $\varphi(t_w)= -
  \beta(t_w) \pi/2$.}
\label{mu}
\end{figure}

\subsection{Effective Temperature derived from oscillatory measurements}
\label{effective}

The GER ratio $\hat\Theta(s)$ is now derived from the oscillatory and
Brownian motion measurements. The selected range $s \in [3;60]$
s$^{-1}$ corresponds to the experimental frequencies range $ f \in
[0.5; 10]$ Hz. From the oscillatory measurements, the variations of
$\tilde \mu(\omega)$ are well represented by the analytical form
$\tilde \mu(\omega) = \mu_0 (\omega/\omega_0)^{\beta} \exp{(-i \beta
\pi /2)}$. The Laplace transform $\hat{\mu}(s) = \tilde \mu(\omega=
is)=\mu_0 (s/s_0)^\beta$ is then derived by analytical
continuation. Again, $s_0$ is an arbitrary reference frequency set to
$s_0 =1$ s$^{-1}$. As can be seen in Fig. \ref{brown}, $\langle{\Delta
r^2}(t)\rangle$ can be approximated by a power-law in the range $t \in
[0.015;0.3]$ s. As a consequence, in the corresponding range $s \in
[3;60]$ s$^{-1}$, the Laplace transform of the mean-squared
displacement is well described by a power-law $\langle\widehat{\Delta
r^2}(s)\rangle = a (s/s_0)^{-b}$. Fig. \ref{param} shows the
parameters $a$, $b$, $\mu_0$ and $\beta$ as a function of aging time $t_w$,
averaged over a set of six different realizations, using the time
rescaling described in section \ref{scaling}.

The exponent $b$ decreases from about $1.9 \pm 0.05$ to $1 \pm 0.05$
while the exponent $\beta$ increases from about zero at low aging
times to $0.82 \pm 0.1$ at the end of the experiment. This is
consistent with the change from a viscous dominated to an elastic
dominated behavior. The
error bars on $\beta$ take into account the departure between the
measured and extrapolated values, which are not systematic (see
another realization in \cite{prl93}).
\begin{figure}
\includegraphics[width=8cm,clip]{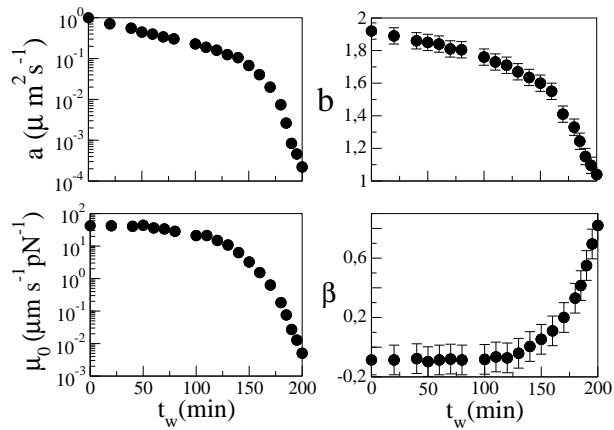}
\caption{\label{param}Top: Brownian motion parameters
  $a$ and $b$ as a function of aging time, where
  $\langle\widehat{\Delta r^2}(s)\rangle = a (s/s_0)^{-b}$, with
  $s_0=1$ s$^{-1}$; Bottom : Mobility parameters
  $\mu_0$ and $\beta$ as a function of aging time, where $\tilde
  \mu(\omega) = \mu_0 (\omega/\omega_0)^{\beta} \exp{(-i \beta \pi
  /2)}$, with $\omega_0 =1$ rad.s$^{-1}$. The results have been
  averaged over six realizations, and rescaled in aging time.}
\end{figure}

Bringing
together our data for $\langle\widehat{\Delta r^2}(s)\rangle$ and
$\hat{\mu}(s)$, and using the modified Generalized Einstein Relation
(\ref{modGER}), the GER ratio $\hat{\Theta}(s)$, parametrized by $t_w$,
is given by : $$\hat{\Theta}(s)= \frac{a}{4 k \mu_0} (s/s_0)^{2-b-\beta}$$
in the range $s \in [3,60]$ s$^{-1}$.

After analytical continuation of $\tilde \Theta (\omega)= \hat{\Theta}(s= -
i\omega)$, we derive the effective temperature $T_{\mbox{\scriptsize eff}}(\omega)$, at
a given $t_w$, from :
\begin{equation}
\label{Teff}
T_{\mbox{\scriptsize eff}} (\omega)=\frac{a}{4 k \mu_0}
\frac{\cos[(b-2)\pi/2]}{\cos(\beta \pi/2)}
(\omega/\omega_0)^{2-b-\beta}
\end{equation}
in the range $\omega \in [3;60]$
rad.s$^{-1}$. In a system at equilibrium, we should have $\frac{a}{4 k \mu_0} = T$ and $2-b -\beta= 0$, which leads to $\hat{\Theta}(s)= T $ and $T_{\mbox{\scriptsize eff}}(\omega) = T$. This is not
the case here for the aging colloidal glass.

The dependence of the effective temperature
$T_{\mbox{\scriptsize eff} }$ with aging time $t_w$, at different frequencies, is shown in
Fig. \ref{T-tw}. The results have been averaged over six
realizations. At the earliest $t_w$, the effective temperature is
close to the bath temperature $T = 300$
K. Upon increase on $t_w$, $T_{\mbox{\scriptsize eff}}$ increases up
to 2-3 times the bath temperature and then decreases back upon further increase on $t_w$. Such a
behavior -- an increase of $T_{\mbox{\scriptsize eff}}$ followed by a
decrease -- was observed for the first time in a colloidal glassy system in
\cite{prl93}, where a scenario for this non-monotonic
behavior was suggested. In the range $\omega \in [3;60]$
rad. s$^{-1}$, $T_{\mbox{\scriptsize eff}}$ increases with frequency for
all aging times as seen in Fig. \ref{T-tw}. In these experiments, the set-up does not allow to
measure $T_{\mbox{\scriptsize eff}}$ at longer $t_w$. Indeed, beyond
$t_w = 200$ minutes, the mobility modulus becomes smaller than
$10^{-2}$ m s$^{-1 }$pN$^{-1 }$. In this range, the optical tweezer is
not powerful enough to induce a detectable motion of the bead. 

Here we remain in the frame of power-law approximation. This is
obviously not true in the full time range, but at least in the range $
[0.015; 0.3]$ s. Significant deviations from a power-law are measured
at large aging times and are currently under study. 

\begin{figure}
\includegraphics[width=8.7cm,clip]{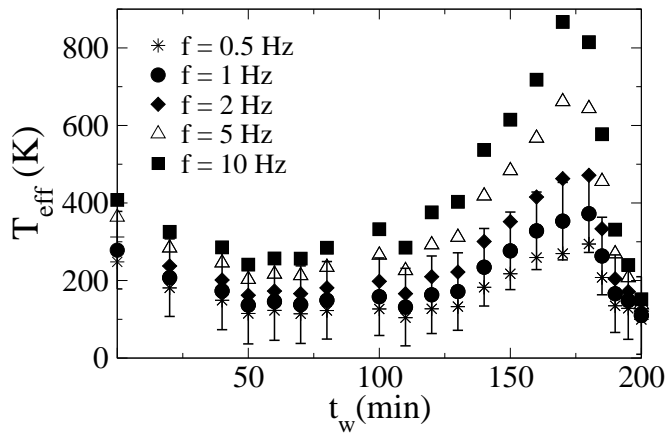}
\caption{\label{T-tw}Effective temperature $T_{\mbox{\scriptsize eff}
  }$ of the colloidal glass as a function of aging time $t_w$,
  measured at different frequencies. Upon increase on $t_w$, $T_{\mbox{\scriptsize
  eff}}$ increases up to about 2-3 times the bath temperature and then
  decreases back. The results have been
  averaged over six realizations rescaled in aging time. The typical errors bars on $T_{eff}$ have only been shown on the data at $f=
1$ Hz for clarity, but all error bars roughly have the same magnitude.}
\end{figure}

\section{Discussion and conclusion}
\label{discussion}

As previously reported
in \cite{prl93}, the dependence of
$T_{\mbox{\scriptsize eff}}$ on $t_w$ can be interpreted through the
change in the glass relaxation times with aging time. In
colloidal glasses, dynamic light scattering and diffusive wave
spectroscopy experiments provide the opportunity to probe the
relaxation times associated to the colloidal particles
diffusion \cite{aboupre64,lequeux,bellour}. Fig. \ref{tau} shows a
scheme of the resulting
relaxation times distribution function $P(\tau)$, centred around
two characteristic relaxation times $\tau_{\rm fast}$ and $\tau_{\rm slow}$. Upon increasing $t_w$, part of the modes distribution
function, centred around $\tau_{\rm slow}$, shifts to larger times,
while the mode at $\tau_{\rm fast} \sim 0.1 $ ms
remains unchanged.
\begin{figure}[h]
\includegraphics[width=9cm]{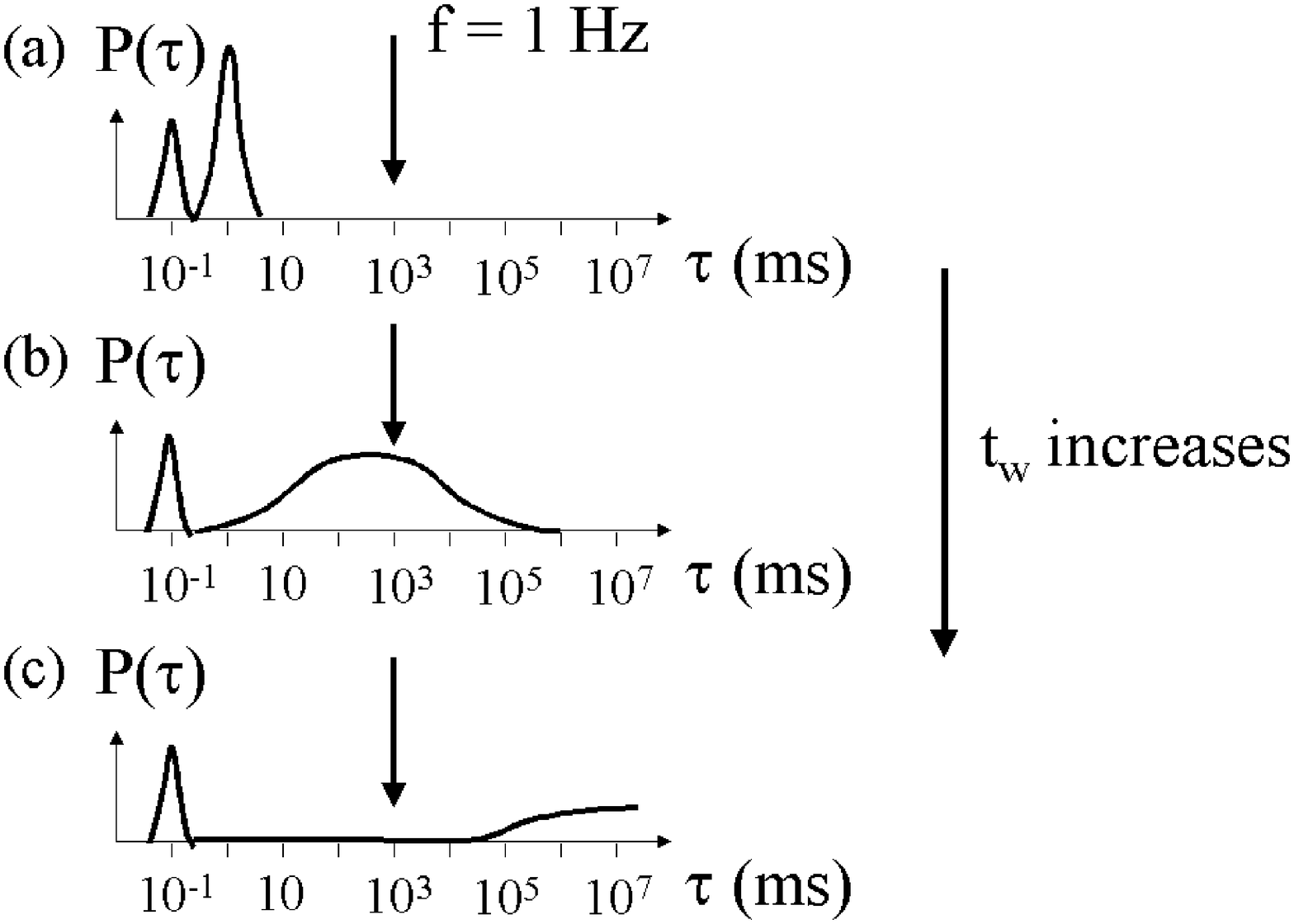}
\caption{ \label{tau} Scheme of the relaxation times distribution function $P(\tau)$ in the glass of Laponite (typically $2.5
\%$ wt) at different aging times $t_w$. Upon increasing $t_w$, part of the modes
distribution, centred around $\tau_{\rm slow}$, shifts towards larger
times, while the mode at $\tau_{\rm fast} \sim
0.1$ ms remains unchanged. The arrow represents the measurements
timescale $1/f$. }
\end{figure}

We suggest the following scenario for the dependence of the
effective temperature on the aging time. When probing the colloidal
glass at a typical frequency $f = 1$ Hz, three situations successively
occur upon increasing $t_w$. At the earliest $t_w$, the glass relaxation
times are small compared to the observation time scale
$\tau_{\rm fast,slow} \ll 1/f $ and do not play any role on this
observation time. The glass is ``{\em at equilibrium}''. In this
experiment, the bead is small enough to be sensitive to molecular
fluctuations. This ensures that it can thermalize with the bath when
the glass relaxation times do not play any role on the
observation time scale. In this case, the reading of the bead
thermometer reduces to the bath temperature $T$. Upon increase on $t_w$, relaxation modes of the order of
the observation time scale $\tau_{\rm slow} \sim 1/ f$ appear in the
glass as seen in Fig. \ref{tau} b. On this observation timescale
$1/f$, the system is {\em out of equilibrium} : the measured
temperature becomes different from the bath temperature. Deviations
from the Generalized Einstein Relation are thus observed. Finally, for
very long $t_w$, the first situation is recovered. The slow and fast
relaxation processes of the glass do not play any role at the
experimental observation timescale because $\tau_{\rm slow} \gg 1/ f$
and $\tau_{\rm fast} \ll 1/ f$. The bead thermometer, only sensitive to the
fast relaxation processes at the molecular scale, again thermalizes with the
bath. The measured temperature is then expected to reduce back to the
bath temperature $T$. Finally, in this scenario, deviations
are only observed in the regime where the observation time scale is of the
order of a characteristic relaxation time of the glass, $\omega
\tau \sim 1$ \cite{prl93}. These different situations are clearly identified in
Fig. \ref{T-tw}. 

As described in \cite{cug97,cug2002}, the temperature of an object is
measured by coupling it to a thermometer during a sufficiently long
time interval such that all heat exchanges between thermometer and
system take place and the whole system equilibrates. In \cite{cug97},
the authors consider an oscillator of characteristic frequency
$\omega_0$ as a thermometer, weakly coupled to a glassy system in
which a regime with small energy flows exist. The glass exhibits
two time correlation scales, a fast one and a slow one, such that:
$R_{ST}(t-t_w)= \frac{1 }{kT}\partial_{t_w} C_{ST}(t-t_w)$ is
satisfied at short times $t-t_w \ll t_w$ and $R_{AG}(t_w /t)= \frac{1
}{kT^*}\partial_{t_w} C_{AG}(t_w /t)$ is satisfied if $t_w /t = O(1)
$. $R_{ST}$ and $C_{ST}$ (or $R_{AG}$ and $C_{AG}$) are respectively
the stationary or aging parts of the response and correlation
functions. $T$ is the bath temperature and $T^{*}$ is the temperature
associated to FDR when $t_w /t = O(1)$.  

If $\omega_0$ is high
enough so that the thermometer evaluates the fast relaxation and
quasiequilibrium is achieved on the time scale $1/\omega_0$, its
asymptotic internal energy density reads $E_{\rm THERM}= k T$. The
reading of the thermometer is the bath temperature $T$. If $\omega_0$
is very low, the thermometer examines the system in its long time
scales behavior and one finds $E_{\rm THERM}= k T^{*}$, where $T^{*}$
is identified as the temperature of the system on this observation
time. Any small but macroscopic thermometer, weakly coupled to the
system, is shown to play the same role as the above considered oscillator. The role of
$\omega_0$ is then played by the inverse of the typical response time
of the thermometer.

In our experiments, the bead plays the role of a thermometer which
characteristic frequency $\omega_0$ is set to the probing frequency
$\omega$. Indeed, the bead inertia is negligible below $\omega_{\rm
inertia} = 10^{8}$ rad.s$^{-1}$ for the highest value of the fluid
viscosity ($\simeq 10$ Pa.s). Experimentally, the observation time
scale $1/ \omega$ is much higher than the typical response time of the
bead thermometer $1/\omega_{\rm inertia}$. As a result, the response
time of the bead is set to the observation time scale $1/\omega$. It
follows that by changing this observation time scale, the thermometer
can explore or be sensitive to, different relaxation time scales of
the glass (see Fig. \ref{tau}). 

In models and simulations -- where the characteristic relaxation time
of the system is set to the waiting time -- deviations are observed
when the characteristic observation time $1/\omega $ is of the same
order or greater than the aging time $t_w$. Experimentally, deviations
are observed in the other regime when $1/\omega \ll t_w$. In our
sense, this discrepancy arises because the aging time $t_w$ is not the only relevant
parameter to describe aging, and that microscopic processes,
characterized by the distribution of relaxation times in the system,
must be considered. In our experiments, deviations are observed when
$\omega \tau \sim 1$. This suggests that, besides the waiting
time $t_w$, the distribution of relaxation times must be included in
models to get an accurate description of aging.

In the past, FDR have been experimentally investigated in the {\em
same colloidal glass} with electrical and rheological measurements
\cite{bellon2}. In the dielectric measurements, the effective
temperature decreases with $t_w$ and $\omega$, and reaches
the bath temperature at high frequency and aging times. In the
rheological measurements, no deviation from FDR could be
detected. More recently, FDR, in the form of a Generalized
Stokes-Einstein relation, has been investigated by combining dynamic
light scattering and rheological measurements \cite{strachan}. The
authors assume that the bulk stress relaxation probed by a rheometer
is the same as the local stress relaxation affecting the probe
particles. The effective temperature is shown to increase with
frequency and aging time in the range $t_w \in [82;135]$ min and
$\omega \in [20;100]$ rad. s$^{-1}$ for a 3$\%$ wt suspension. The
authors suggest the existence of two regimes: $\omega \tau \ll 1$
leading to the bath temperature $T$ and $\omega \tau \gg 1$ leading to $T_{\rm
eff}$, which is not consistent with our interpretation despite the
authors claim. Other groups have recently investigated the deviations from FDR,
by performing a test of the Stokes-Einstein relation, using optical
tweezers \cite{jabbari,greinert}. In one case, they found that no deviation
could be detected in the aging colloidal glass over a wide range of
frequencies $f \in [1.2;12000]$ Hz. In the other one, deviations are
observed at large aging times. Although the reason of these
discrepancies have to be elucidated, our experimental findings are
fully supported by our interpretation based on microscopic processes
probed in light scattering experiments (see Fig. \ref{tau}).

At this stage, whether these experimental investigations with the same
observables, in the same colloidal glass, are in contradiction still
remains an open question \cite{prl93,strachan,jabbari}. Furthermore,
the definition of an effective temperature, with a thermodynamic
meaning, implies that this quantity is independent of the chosen
observable, which is not verified in \cite{bellon2} where
rheological and dielectric measurements do not lead to the same
effective temperature. Is the
effective temperature defined through a generalisation of FDR a
relevant concept ? Does it depend on the chosen observable, as shown
in the glass phase of Bouchaud's trap model \cite{fielding}, or not,
as in simulations on a binary Lennard-Jones mixture
\cite{berthier} ? More experimental investigations are needed to
understand these contradictory findings (see \cite{cip-ramos} for a review). 

In conclusion, this work provides an experimental and theoretical
investigation of the Generalized Einstein Relation in an aging
colloidal glass. We interpret the observed deviations from GER as
directly related to the change in the glass relaxation
times. Deviations are observed in the regime $\omega \tau \sim 1$.

As a final comment, let us add that dynamical heterogeneities are recognized as a general feature of slow
dynamics encountered in supercooled liquids and glasses
\cite{ediger,weeks,duri}. In supercooled liquids, deviations from the
Einstein relation were explained by such dynamical heterogeneities
\cite{ediger}. Up to now, in models and simulations, deviations from
FDR have been found in homogeneous glassy systems and interpreted in
terms of an effective temperature. Moreover, the effective temperature is
defined as an ensemble-averaged quantity in experiments, models and
simulations. Understanding how the concept of effective temperature
interplays with such heterogeneities is an open question.


\begin{thebibliography}{00}

\bibitem{cug93} L.F. Cugliandolo and J. Kurchan, {\it
Phys. Rev. Lett.} {\bf 71}, 173 -176 (1993).
 
\bibitem{cug97}
L.F Cugliandolo, J. Kurchan, and L. Peliti, {\it Phys. Rev. E} {\bf 55}, 3898 - 3914 (1997).  

\bibitem{parisi}
G. Parisi, {\it Phys. Rev. Lett.} {\bf 79}, 3660 - 3663 (1997).

\bibitem{barrat}
A. Barrat, {\it Phys. Rev. E } {\bf 57}, 3629 - 3632 (1998).   

\bibitem{sellitto}
M. Sellitto, {\it Eur. Phys. Journ. B} {\bf 4}, 135 - 138 (1998).

\bibitem{marinari}
E. Marinari, G. Parisi, F. Ricci-Tersenghi, and J.J. Ruiz-Lorenzo, {\it J. Phys. A : Math Gen.} {\bf 31}, 2611 - 2620 (1998).

\bibitem{barrat2}
J.-L. Barrat and W. Kob, {\it Europhys. Lett.} {\bf 46}, 637 - 642 (1999).

\bibitem{barrat3}
L. Berthier, J.-L. Barrat, and J. Kurchan, {\it Phys. Rev. E} {\bf 61}, 5464 -
5472 (2000).

\bibitem{makse}
H.A. Makse and J. Kurchan, {\it Nature} {\bf 415}, 614 - 617 (2002).  

\bibitem{fielding}
S. Fielding and P. Sollich, {\it Phys. Rev. Lett.} {\bf 88}, 050603-1 -
050603-4 (2002).

\bibitem{grigera} T.S. Grigera and N.E. Israeloff, {\it Phys. Rev. Lett.} {\bf 83}, 5038 - 5041 (1999).

\bibitem{bellon}
L. Bellon, S. Ciliberto, and C. Laroche, {\it Europhys. Lett.} {\bf
  53}, 511 - 517 (2001).

\bibitem{bellon2}
L. Bellon and S. Ciliberto, {\it Physica D} {\bf
  168 - 169 }, 325 - 335 (2002). 

\bibitem{prl93} B. Abou and F. Gallet, {\it Phys. Rev. Lett.} {\bf
  93}, 160603 1-4 (2004).

\bibitem{strachan}
D.R. Strachan, G.C. Kalur, and S.R. Raghavan, {\it Phys. Rev. E} {\bf
  73}, 041509 1-5 (2006). 

\bibitem{jabbari} S. Jabbari-Farouji, D. Mizuno, M. Atakhorrami,
F.C. MacKintosh, C.F. Schmidt, E. Eiser, G.H. Wegdam, and D. Bonn,
{\it Phys. Rev. Lett.} {\bf 98}, 108302 (2007).

\bibitem{greinert}
N. Greinert, T. Wood, and P. Bartlett, {\it Phys. Rev. Lett.} {\bf 97}, 257202 (2006).

\bibitem{herisson} D. H\'erisson and M. Ocio, {\it Phys. Rev. Lett.} {\bf 88}, 257202
- 257205 (2002).
 
\bibitem{buisson} L. Buisson, S. Ciliberto, and A. Garcimart\'{\i}n,
  {\it Europhys. Lett.} {\bf 63}, 603 - 609 (2003).

\bibitem{danna}
G. D'Anna, P. Mayor, A. Barrat, V. Loreto, and F. Nori, {\it Nature} {\bf 424},
1909 - 912 (2003). 

\bibitem{oja}
R.P. Ojha, P.-A. Lemieux, P.K. Dixon, A.J. Liu, and D.J. Durian, {\it Nature} {\bf 427}, 521 - 523
(2004). 

\bibitem{Kubo} R. Kubo, {\it Rep. Prog. Phys.} {\bf 29}, 255 - 284
(1966); R. Kubo, M. Toda, and N. Hashitsume, Statistical physics
\uppercase\expandafter{\romannumeral 2} : nonequilibrium statistical
mechanics, Second edition, Springer-Verlag, Berlin, 1991.

\bibitem{pass-micro} T.G. Mason, K. Ganesan, J.H. van Zanten,
  D. Wirtz, and S.C. Kuo, {\it Phys. Rev. Lett.} {\bf 79}, 3282 -
  3285 (1997).

\bibitem{pottier2003} N. Pottier and A. Mauger, {\it Physica A} {\bf 332}, 15 - 28 (2004).

\bibitem{pottier2005} N. Pottier {\it Physica A} {\bf
  345}, 472 - 484 (2005).

\bibitem{Kroon} M. Kroon, W.L. Vos, and G.H. Wegdam, {\it Phys. Rev. E} {\bf
57}, 1962-1970 (1998).

\bibitem{Thompson and Butterworth} D.W. Thompson and J.T. Butterworth, {\it J. Colloid Interface Sci.} {\bf 151}, 236-243 (1992).

\bibitem{wigner} D. Bonn, S. Tanaka, G.H. Wegdam, H. Kellay, and
J. Meunier, {\it Europhys. Lett.}
{\bf 45}, 52 - 57 1(1999).

\bibitem{aboupre64}
B. Abou, D. Bonn, and J. Meunier, {\it Phys. Rev. E} {\bf 64},
021510 - 021513 (2001). 


\bibitem{jorlapo} B. Abou, D. Bonn, and J. Meunier, {\it
J. Rheol.} {\bf 47}, 979 - 988 (2003).

\bibitem{henon} S. H\'enon, G. Lenormand, A. Richert, and F. Gallet,
{\it Biophys. J. } {\bf 76}, 1145 - 1151 (1999).

\bibitem{joshi} Y. Joshi, cond-mat/0702467 (2007).

\bibitem{prl89} D. Bonn, S. Tanase, B. Abou, H. Tanaka and J. Meunier,
  {\it Phys. Rev. Lett.} {\bf 89}, 015701-1 (2002). 

\bibitem{balland} M. Balland, A. Richert, and F. Gallet,
{\it Eur. Biophys. J. } {\bf 34}, 255 - 261 (2005).

\bibitem{berret} J.-F. Berret, {\it Molecular gels}, Springer-Verlag,
  Weiss (Eds), 2005; cond-mat/0406681. 
 
\bibitem{lequeux}
A. Knaebel, M. Bellour, J.-P. Munch, V. Viasnoff, F. Lequeux, and
J.L. Harden, {\it Europhys. Lett.} {\bf
  52}, 73 - 79 (2000). 

\bibitem{bellour}
M. Bellour, A. Knaebel, J.L. Harden, F. Lequeux, and J.-P. Munch, {\it Phys. Rev. E} {\bf 67},
031405 1-8 (2003). 

\bibitem{cug2002}
L.F. Cugliandolo, {\it Lecture notes, Les Houches} (2002); cond-mat/0210312. 

\bibitem{berthier}
L. Berthier and J.L. Barrat, {\it J. Chem. Phys.} {\bf
  116}, 6228 - 6242 (2002). 

\bibitem{cip-ramos}
L. Cipelletti and L. Ramos, {\it Journal of Physics: Cond. Matt.}
{\bf 17}, R253 - R285 (2005).

\bibitem{ediger} S.F. Swallen, P.A. Bonvallet, R.J. McMahon, and
M.D. Ediger, {\it Phys. Rev. Lett.} {\bf 90}, 015901 1-4 (2003).

\bibitem{weeks}
E.R. Weeks, J.C. Crocker, A.C. Levitt, A. Schofield, and D.A. Weitz, {\it Science} {\bf 287}, 627 - 631 (2000).

\bibitem{duri} A. Duri, P. Ballesta, L. Cipelletti, H. Bissig, and
V. Trappe, {\it Fluctuation and noise Letters} {\bf 5}, L1 (2005).


\end{thebibliography}
\end{document}